\documentstyle[aaspp4,flushrt]{article}

\catcode`\@=11 
\def\@versim#1#2{\vcenter{\offinterlineskip
        \ialign{$\m@th#1\hfil##\hfil$\crcr#2\crcr\sim\crcr } }}
\newcommand{\Ref}{\hangindent=20pt \hangafter=1 \noindent}
\newcommand{\StartRef}{\hyphenpenalty=10000 \raggedright}

\newcommand{\beq}{\begin{equation}}
\newcommand{\eeq}{\end{equation}}
\def\lsim{\mathrel{\mathpalette\@versim<}}
\def\gsim{\mathrel{\mathpalette\@versim>}}

\def\@versim#1#2{\vcenter{\offinterlineskip
        \ialign{$\m@th#1\hfil##\hfil$\crcr#2\crcr\sim\crcr } }}
\catcode`\@=12 
\begin{document}
\title{Thermal X-Ray Line Emission from Accreting Black Holes}
\author{Ramesh Narayan\footnote{rnarayan@cfa.harvard.edu} and 
John Raymond\footnote{jraymond@cfa.harvard.edu}} 
\affil{Harvard-Smithsonian Center for Astrophysics, 60 Garden St., 
Cambridge, MA 02138}
\medskip
\setcounter{footnote}{0}

\begin{abstract}
We present model X-ray spectra of accreting black holes with
advection-dominated accretion flows, paying attention to thermal
emission lines from the hot plasma.  We show that the Advanced X-ray
Astrophysical Facility (AXAF) might be able to observe lines from
X-ray binaries such as V404 Cyg in quiescence, the Galactic center
black hole Sagittarius A$^*$, and the nuclei of nearby galaxies such
as M87.  Line intensities can provide new diagnostics to study the
accreting plasma in these and related systems.

\noindent {\em Subject headings:} Accretion, Black Holes, Galaxies:
Nuclei, Galaxy: Center, Radiation Mechanisms: Thermal, X-Rays:
Binaries, X-Rays: Spectra
\end{abstract}

\section{Introduction}

The AXAF satellite will enable us to measure X-ray spectra of
astronomical sources with good spectral resolution --- about 60 eV at
1 keV for the AXAF CCD Imaging Spectrometer, ACIS --- and excellent
spatial resolution (better than 1 arcsecond).  With an effective area
of $\sim300$ cm$^2$, ACIS should observe spectral lines from a number
of X-ray sources.  The XMM satellite may provide even greater
sensitivity for lines.  Measurements of line intensities will provide
important new diagnostics for the physical conditions of the radiating
plasma.

Some of the most important X-ray sources in the sky are accreting
black holes.  Examples are stellar mass black holes in X-ray binaries,
and supermassive black holes in the nucleus of our Galaxy and other
galaxies.  X-ray emission from these sources arises as a result of
optically thick thermal radiation from an accretion disk (cf. Tanaka
\& Shibazaki 1996), or by Compton-scattering of soft photons by a hot
optically thin plasma, either in a corona (e.g. Haardt \& Maraschi
1991) or in an advection-dominated accretion flow (ADAF, Narayan \& Yi
1994, 1995, Abramowicz et al. 1995, see Narayan, Mahadevan \& Quataert
1998b and Kato, Fukue \& Mineshige 1998 for reviews), or by thermal
bremsstrahlung from hot low density gas.  The spectral energy
distributions corresponding to these different processes have been
studied by a number of authors (see Liang 1998 and Mushotzky, Done \&
Pounds 1993 for reviews).

Fewer models are available for X-ray line emission.  Fluorescent line
emission from cool gas irradiated by a hot corona has been studied in
some detail (Mushotzky et al. 1993, Tanaka et al. 1995), but thermal
X-ray line emission from hot optically thin gas around black holes has
not been discussed very much.  In the present Letter, we show that
black holes that accrete via ADAFs with relatively low mass accretion
rates might produce X-ray lines detectable with AXAF or XMM.  These
lines could provide new constraints on the accretion flows in these
systems.  We point out three significant diagnostic possibilities.
First, the equivalent widths of the emission lines increase with the
size of the ADAF region.  Second, for a given ADAF size the equivalent
widths are much larger for models that include a wind than for models
with no wind. And, third, photoionization is unimportant in an ADAF,
while it dominates in some accretion disk corona models.

\section{ADAF Models}

Both corona models and ADAF models feature hot optically thin gas, and
both lead to thermal X-ray emission.  However, corona models are
usually very Compton-dominated, so that any thermal lines are
likely to have low equivalent widths.  ADAFs too are Compton-dominated
when the mass accretion rate is high; if the Eddington-scaled
accretion rate, $\dot m\equiv \dot M/2.2\times10^{-8}m ~M_\odot {\rm
yr^{-1}}$ (where $m$ is the black hole mass in solar units), is close
to a critical value $\dot m_{crit}\sim0.05-0.1$ (Esin, McClintock \&
Narayan 1997), the Comptonization of soft photons (either from an
outer disk or from thermal synchrotron emission) is so strong that it
greatly exceeds the thermal emission.

As the mass accretion rate in an ADAF decreases to $\dot m\lsim 0.01$,
thermal bremsstrahlung becomes relatively more important (Esin et
al. 1997); thermal lines should then have measureable equivalent
widths.  The electron temperature in an ADAF varies roughly as
$T_e\sim10^{12}K/r$ for $r\gsim10^3$ (Narayan \& Yi 1995), where $r$
is the radius in Schwarzschild units ($2.95\times10^5m$ cm).  Since
the strongest X-ray line emission arises from thermal gas with
$T_e\sim10^7-10^8K$, the region of the flow between $r\sim10^4-10^5$
is most important for our purpose.  ADAFs can extend to such large
radii only for low $\dot m$ (Narayan \& Yi 1995, Narayan et
al. 1998b), another reason to concentrate on low-$\dot m$ systems.

We have computed ADAF models and corresponding X-ray spectra of the
following three low-$\dot m$ systems: the X-ray binary V404 Cyg in
quiescence, the Galactic Center source Sgr A$^*$, and the nucleus of
the galaxy M87.  The methods are described in Quataert \& Narayan
(1998, and references therein).  ADAF models are not unique, as there
is considerable uncertainty associated with possible winds (Narayan \&
Yi 1994, Blandford \& Begelman 1998, Quataert \& Narayan 1998).
Therefore, for each of our three systems we have computed two models,
one in which there is no wind (referred to as NW, the accretion rate
$\dot m$ is taken to be independent of radius), and one with a
moderately strong wind (referred to as W, $\dot m$ is assumed to vary
with radius as $r^{0.4}$).  These two models probably bracket the true
situation.  We assume that in the NW model the electrons in the
accreting plasma receive only a fraction $\delta=0.01$ of the
viscously dissipated heat in the accretion flow.  However, in the W
model we set $\delta=0.3$, as recommended by Quataert \& Narayan
(1998), in order to fit the continuum spectrum.  We assign reasonable
values to the other microscopic parameters: viscosity parameter
$\alpha=0.1$, plasma $\beta=10$.  The results are not very sensitive
to the values of these parameters.

The system-specific parameters are as follows: see Narayan, Barret \&
McClintock (1997) for a compilation of the data on V404 Cyg, Narayan
et al. (1998a) for Sgr A$^*$, and Reynolds et al. (1996) for M87.  The
black hole masses are $m=12$ (V404 Cyg), $m=2.5\times10^6$ (Sgr
A$^*$), $m=3\times10^9$ (M87), the distances are 3.5 kpc (V404 Cyg),
8.5 kpc (Sgr A$^*$), 16 Mpc (M87), and the absorbing columns are
$N_H=1.1\times10^{22} ~{\rm cm^{-2}}$ (V404 Cyg), $N_H=6\times10^{22}
~{\rm cm^{-2}}$ (Sgr A$^*$), $N_H=2.5\times10^{20} ~{\rm cm^{-2}}$
(M87, this $N_H$ includes only the contribution of our Galaxy).  The
mass accretion rates are not known a priori.  We fit $\dot m$ in each
model so as to reproduce the observed X-ray continuum.  This gives
$\dot m =$ 0.0010, 0.0060 (NW and W models of V404 Cyg), 0.000070,
0.00024 (Sgr A$^*$), 0.0010, 0.011 (M87).  For the W models, where
$\dot m$ varies with $r$, the quoted $\dot m$ corresponds to the outer
edge of the ADAF.

The greatest uncertainty in the models has to do with the size of the
ADAF region.  In V404 Cyg, the ADAF is believed to extend from the
black hole horizon ($r=1$) to a transition radius $r_{tr}$, beyond
which the flow consists of a thin disk plus a corona.  We take
$r_{tr}=10^{4.4}$, as indicated by the width of the $H_\alpha$ line
from the outer disk (Narayan et al. 1997).  In the coronal region, we
set $\dot m\propto r_{tr}/r$ to model the evaporation of gas from the
disk into the corona (cf. Esin et al. 1997).  In Sgr A$^*$, there is
no evidence for an outer disk, so we consider a pure ADAF model
extending from $r=1$ to an outer radius $r_{out}$.  The stars whose
winds supply most of the accreting mass (cf. Coker \& Melia 1997) are
located at $r\gsim {\rm few}\times10^5$.  We therefore set
$r_{out}=10^5$.  For simplicity, we use the same value of $r_{out}$
for M87 as well.

\section{Model X-Ray Line Spectra}

Each quasi-spherical ADAF model gives the electron density $n_e$ and
electron temperature $T_e$ as a function of $r$ on a radial grid with
ten zones per decade of $r$.  Figure 1 shows these quantities for the
NW and W models of Sgr A$^*$.

For calculating the line spectrum, we divide the flow into an inner
region from $r=1-10^2$ and an outer region beyond $10^2$.  We compute
the continuum emission from the inner region due to synchrotron,
bremsstrahlung and Comptonization, but we do not compute line
emission; this is reasonable since the temperature is $>10^9$ K
(Fig. 1), and the astrophysically abundant elements are fully ionized.
In the outer region, we compute the line emission and thermal
continuum in detail, but we do not consider either synchrotron
emission or the effect of Comptonization; these latter processes are
very steep functions of the temperature, and are negligible outside
$r=10^2$.  The spectra we present include only the radiation from the
accreting gas; we ignore any emission by the wind.

The spectral calculation in the outer region employs an updated
version of the Raymond \& Smith (1977) X-ray code to compute the
bremsstrahlung, recombination and two-photon continua and the emission
in spectral lines.  For the present models we are mostly concerned
with H-like and He-like ions.  Collisional excitation (e.g. Pradhan,
Norcross \& Hummer 1981; Pradhan 1985), recombination to excited
levels (e.g. Mewe, Schrijver \& Sylwester 1980) and dielectronic
recombination satellite lines (e.g. Dubau et al 1981; Bely-Daubau et
al 1982) are included.  The calculation has been done with a spectral
binning of 60 eV over a range of 0.2-30 keV.  We have ignored Doppler
smearing of the lines, which for an ADAF is of order the thermal
broadening, and is relatively unimportant at the large radii of
interest.

The calculations proceed outward from the innermost zone at $r=10^2$.
In each zone, we use the local density and temperature to compute the
thermal emission.  We include photoionization by all radiation emitted
at smaller radii (including the continuum emission from the region
inside $r=10^2$).  However, we find that photoionization affects the
ionization state of the abundant ions by at most a few per cent.  This
results from the relatively inefficient conversion of accretion energy
to X-ray luminosity in an ADAF ($< 0.1\%$ in the models presented
here).  It stands in contrast to the dominance of photoionization in
the coronae of Low Mass X-ray Binaries, which convert $\sim 20\%$ of
the accretion energy to X-rays (Raymond 1993; Ko \& Kallman 1994).  We
have not included photoabsorption, but we confirmed a posteriori that
it is not important in reducing the X-ray flux.

\section{Results}

Figure 2 shows model X-ray spectra of V404 Cyg, Sgr A$^*$ and M87, for
NW models (upper panel) and W models (lower panel); the spectra have
been corrected for absorption using the estimated $N_H$.  The ordinate
shows the photon count rate per 60 eV bin in units of ${\rm
cm^{-2}\,s^{-1}}$.  For the ACIS detector, with an effective area of
$\sim300$ cm$^2$, and an integration time of $3\times10^4$ s, the
number of counts we may expect to detect in an observation is the
photon count rate multiplied by $\sim10^7~{\rm cm^2\,s}$.  We see from
Fig. 2 that several of the lines in the model spectra could be
observed with good signal to noise in such an observation.

Table 1 lists equivalent widths of a few selected lines. In general,
we expect to see lines of H-like and He-like ions of high-Z elements.
Figure 3 shows the equivalent widths of a few lines as a function of
outer radius for the Sgr A* model with a wind.  The equivalent width
of each line rises as the temperature declines to the characteristic
temperature at which the line is formed.  As the temperature falls
still lower, there is little emission in either the line or the
continuum at that energy, and the equivalent width remains constant.
Most lines form at fairly large radii, $r\sim10^4-10^5$.  Clearly, by
studying the relative intensities of different lines, it would be
possible to estimate how much gas is present at different
temperatures; moreover, because of the near one-to-one mapping between
temperature and radius at the radii of interest (see Fig. 1), we could
estimate the run of gas density with radius.

Both Fig. 2 and Table 1 show a striking difference between non-wind
(NW) and wind (W) models, with the latter producing significantly
stronger lines.  NW models are typically more Compton-dominated than W
models (Quataert \& Narayan 1998), and therefore have less thermal
emission.  Also, because $\dot m$ is a rising function of $r$ ($\dot
m\propto r^{0.4}$), W models have relatively more gas at larger radii
(see Fig. 1).  Since the bulk of the line emission comes from radii
outside about $10^4$ (Fig. 3), this dramatically enhances the line
emission in these models.

\section{Discussion}

The main result of this Letter is that high resolution X-ray spectra
of low-luminosity accreting black holes could reveal interesting
emission lines if the accretion in these sources occurs via an ADAF.
Some of the X-ray lines could be observed with good signal to noise
with the ACIS detector on AXAF, and with XMM, ASTRO E or the proposed
Constellation--X.

Detection of X-ray lines with any reasonable strength would imply an
ADAF with a large outer radius, and would rule out a Compton-dominated
corona model.  Also, models with winds have significantly larger
equivalent widths than those without.  As Fig. 3 shows, different
lines arise from different radii.  Therefore, if the strengths of many
lines are measured, both the size of the ADAF and the presence or
absence of a wind could be determined.  In principle, the composition
of the gas could also be estimated.

In addition, the models predict that photoionization is unimportant,
which is a unique feature of the ADAF model and distinguishes it from
corona models.  There are several X-ray diagnostics with which it
should be possible to distinguish collisionally ionized hot plasma
from photoionized gas (Liedahl et al 1992; Kallman 1995).  For the
present case, the most promising diagnostic would be the ratio of the
$\rm 1s^2 - 1s2p ^1P$ resonance line to the $\rm 1s^2 - 1s2s ^3S$
forbidden line in any of the He-like ions (e.g. the $\lambda1.850${\AA}
and $\lambda1.867${\AA} lines of Fe XXV).

We should emphasize that the models presented in this Letter are only
illustrations of typical effects that one may expect in sources with
ADAFs.  In the wind models, for instance, we calculate only the line
emission from the accreting gas and ignore any radiation from the
wind.  In addition, there could be line emission from diffuse hot gas
surrounding the source (e.g. in Sgr A$^*$, Koyama et al. 1996), though
such emission is not a serious problem with an instrument like ACIS
which has excellent angular resolution.

Of the models discussed here, those of V404 Cyg are perhaps the most
secure.  It is the only one of the three sources for which the X-ray
luminosity is known accurately (Narayan et al. 1997), and for which we
have a reliable estimate of $\dot m$.  The size of the ADAF zone is
also reasonably constrained.  However, the model uses a simplified
description of the coronal gas above the outer thin disk.

The situation is worse for the other two sources.  The X-ray emission
from the accretion flow in Sgr A$^*$ is not well constrained, both
because of the poor angular resolution of current observations which
cannot distinguish the black hole from the surrounding diffuse gas
(Koyama et al. 1996), and because of the large uncertainty in the
absorbing column to the source.  Both problems can be solved with ACIS
observations.  If the X-ray continuum is not much below the estimate
used by Narayan et al. (1998a), then Fig. 2 indicates that line
emission would be relatively easy to observe.

In the case of M87, it is not clear that the observed X-rays (Reynolds
et al. 1996) necessarily come from the accretion flow onto the central
black hole.  The size of the ADAF too is highly uncertain; $r_{out}$
could have any value from $10^2-10^5$.  Observations with AXAF will
clarify the situation.  In an optimistic scenario (lower panel of
Fig. 2), lines should be seen readily.

\noindent{\it Acknowledgments.}  We thank Josh Grindlay for useful
discussions on the capabilities of AXAF and XMM, and Jeff McClintock,
Kristen Menou and Eliot Quataert for comments on the manuscript.  This
work was supported in part by NASA grants NAG 5-2837 and 5-2845 to the
Smithsonian Astrophysical Observatory.

\bigskip\bigskip
{
\footnotesize
\StartRef
\noindent {\large \bf References} \\

\Ref Abramowicz, M., Chen, X., Kato, S., Lasota, J.-P., \& Regev, O., 1995,
ApJ, 438, L37 \\
\Ref Blandford, R. D., \& Begelman, M. C., 1998, MNRAS submitted 
(astro-ph/9809083) \\
\Ref Bely-Daubau, F., Faucher, P., Dubau, J., \& Gabriel, A.H. 1982, 
MNRAS, 198, 239 \\
\Ref Coker, R. F., \& Melia, F. 1997, ApJ, 488, L149 \\
\Ref Dubau, J., Loulergue, M., Gabriel, A.H., Steenman-Clark, L., \& Volonte, 
S. 1981, MNRAS, 195, 705 \\
\Ref Esin, A. A., McClintock, J. E., \& Narayan, R. 1997, ApJ, 489, 867 \\
\Ref Haardt, F., \& Maraschi, L. 1991, ApJ, 380, L51 \\
\Ref Kallman, T.R. 1995, ApJ, 455, 603 \\
\Ref Kato, S., Fukue, J., \& Mineshige, S. 1998, Black-Hole Accretion Disks
(Kyoto: Kyoto Univ. Press) \\
\Ref Ko, Y.-K., \& Kallman, T.R. 1994, ApJ, 431, 273 \\
\Ref Koyama, K. et al. 1996, PASJ, 48, 249 \\
\Ref Liang, E. P. 1998, Phys. Repts., 302, 67 \\
\Ref Liedahl D.A., Kahn, S.M., Osterheld, A.L., \& Goldstein, W.H. 1992, 
ApJ, 391, 306 \\
\Ref Mewe, R., \& Schrijver, J., \& Sylwester, J. 1980, A\& AS, 40, 323 \\
\Ref Mushotzky, R. F., Done, C., \& Pounds, K. A. 1993, ARA\&A, 31, 717 \\
\Ref Narayan, R., Barret, D., \& McClintock, J. E. 1997, ApJ, 482, 448 \\
\Ref Narayan, R., Mahadevan, R., Grindlay, J. E., Popham, R. G., \& Gammie,
C. 1998a, ApJ, 492, 554 \\
\Ref Narayan, R., Mahadevan, R., \& Quataert, E. 1998b, The Theory of Black
Hole Accretion disks, eds. M. A. Abramowicz, G. Bjornsson, \& J. E. Pringle
(Cambridge Univ. Press), in press (astro-ph/9803141) \\
\Ref Narayan, R., \& Yi, I. 1994, ApJ, 428, L13 \\
\Ref Narayan, R., \& Yi, I. 1995, ApJ, 452, 710 \\
\Ref Pradhan, A.K. 1985, ApJ Suppl., 59, 183 \\
\Ref Pradhan, A.K., Norcross, D.W., \& Hummer, D.G. 1981, ApJ, 246, 1031 \\
\Ref Quataert, E., \& Narayan, R. 1998, ApJ, sumitted (astro-ph/9810136) \\
\Ref Raymond, J.C., \& Smith, B.W. 1977, ApJ Suppl., 35, 419 \\
\Ref Raymond, J.C. 1993, ApJ, 412, 267 \\
\Ref Reynolds, C. S., Di Matteo, T., Fabian, A. C., Hwang, U., \&
Canizares, C. R. 1996, MNRAS, 283, L111 \\
\Ref Tanaka, Y., \& Shibazaki, N. 1996, ARA\&A, 34, 607 \\
\Ref Tanaka, Y. et al. 1995, Nature, 375, 659 \\

\newpage

\begin{deluxetable}{lcccccccc}
\tablecaption{Equivalent widths (in eV) of some X-ray lines in ADAF
models of V404 Cyg with No Wind (VNW) and with Wind (VW), Sgr A$^*$
(SNW, SW), and M87 (MNW, MW).}  
\tablewidth{0pt}

\tablehead{ \colhead{Ion} & \colhead{$\lambda$ (\AA)} & \colhead{VNW} &
\colhead{VW} & \colhead{SNW} & \colhead{SW} & \colhead{MNW} &
\colhead{MW} & }

\startdata

O VII & 21.60 & 0.0 & 0.0 & 0.0 & 0.6 & 0.0 & 0.6 \nl
O VIII & 18.97 & 0.0 & 1.3 & 3.1 & 47.5 & 0.2 & 47.2 \nl
Mg XI & 9.168 & 0.0 & 0.1 & 0.3 & 5.6 & 0.0 & 5.4 \nl
Mg XII & 8.425 & 0.0 & 1.4 & 2.6 & 31.0 & 0.2 & 30.6 \nl
Si XIII & 6.648 & 0.0 & 0.8 & 2.4 & 37.4 & 0.2 & 36.3 \nl
Si XIV & 6.180 & 0.1 & 6.0 & 8.2 & 79.8 & 0.6 & 79.1 \nl
S XV & 5.039 & 0.0 & 1.8 & 3.2 & 40.4 & 0.2 & 39.4 \nl
S XVI & 4.727 & 0.1 & 7.2 & 6.7 & 50.1 & 0.5 & 49.8 \nl
Ar XVII & 3.949 & 0.0 & 2.9 & 3.0 & 31.0 & 0.2 & 30.3 \nl
Ar XVIII & 3.731 & 0.1 & 7.1 & 4.8 & 28.4 & 0.4 & 28.3 \nl
Ca XIX & 3.173 & 0.0 & 1.7 & 1.1 & 10.1 & 0.1 & 9.8 \nl
Ca XX & 3.020 & 0.0 & 2.7 & 1.6 & 7.8 & 0.1 & 7.8 \nl
Fe XXIV & 1.861 & 0.0 & 1.6 & 0.7 & 7.5 & 0.1 & 7.3 \nl
Fe XXV & 1.850 & 1.0 & 110 & 50.9 & 327 & 3.6 & 315 \nl
Fe XXV & 1.867 & 0.5 & 41.7 & 19.5 & 124 & 1.6 & 121 \nl
Fe XXV & 1.859 & 0.5 & 43.5 & 19.3 & 134 & 1.5 & 131 \nl
Fe XXV & 1.590 & 0.2 & 23.0 & 10.2 & 70.0 & 0.7 & 67.1 \nl
Fe XXV & 1.855 & 0.2 & 35.0 & 15.5 & 160 & 1.2 & 155 \nl
Fe XXVI & 1.780 & 1.7 & 79.1 & 60.5 & 190 & 4.7 & 189 \nl
Fe XXVI & 1.500 & 0.4 & 16.4 & 13.7 & 40.4 & 1.1 & 40.6 \nl
Ni XXVII & 1.587 & 0.1 & 6.4 & 3.1 & 18.0 & 0.2 & 17.3 \nl
Ni XXVIII & 1.530 & 0.1 & 3.9 & 3.4 & 9.4 & 0.3 & 9.5 \nl

\enddata
\label{tab-v}
\end{deluxetable}

\newpage
\vskip 5in
\newpage

\begin{figure}
\plotone{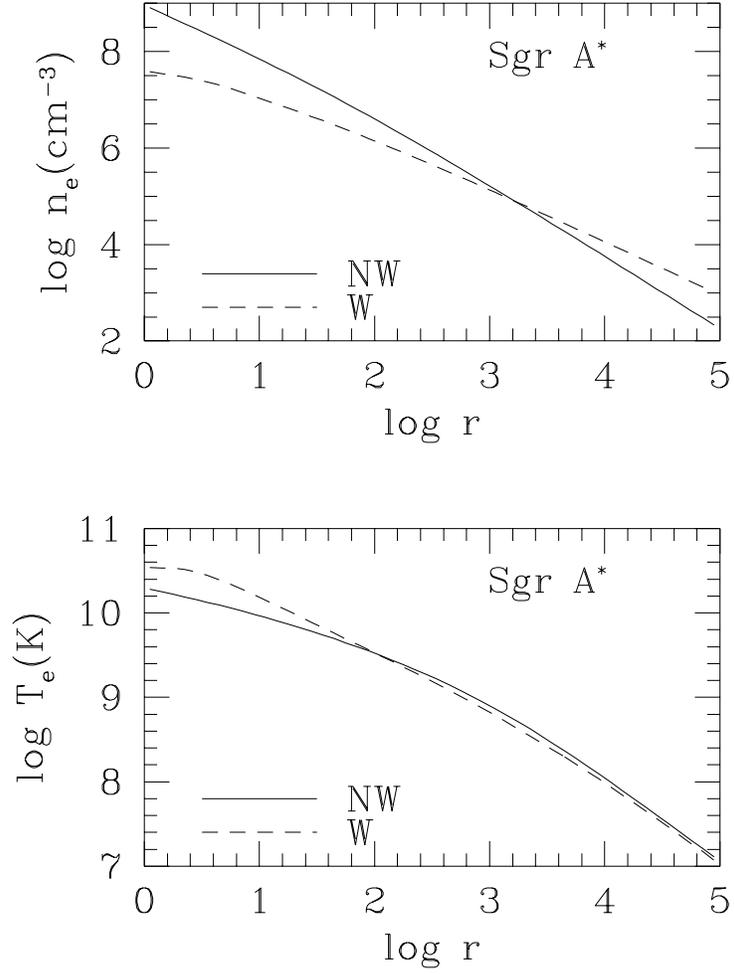}
\caption{Electron density $n_e$ and electron temperature $T_e$ versus
radius $r$ (in Schwarzschild units) for two models of Sgr A$^*$.  NW
refers to a model with no wind ($\dot m$ independent of $r$), and W to
one with a wind ($\dot m\propto r^{0.4}$).}
\end{figure}

\newpage
\vskip 5in
\newpage

\begin{figure}
\plotone{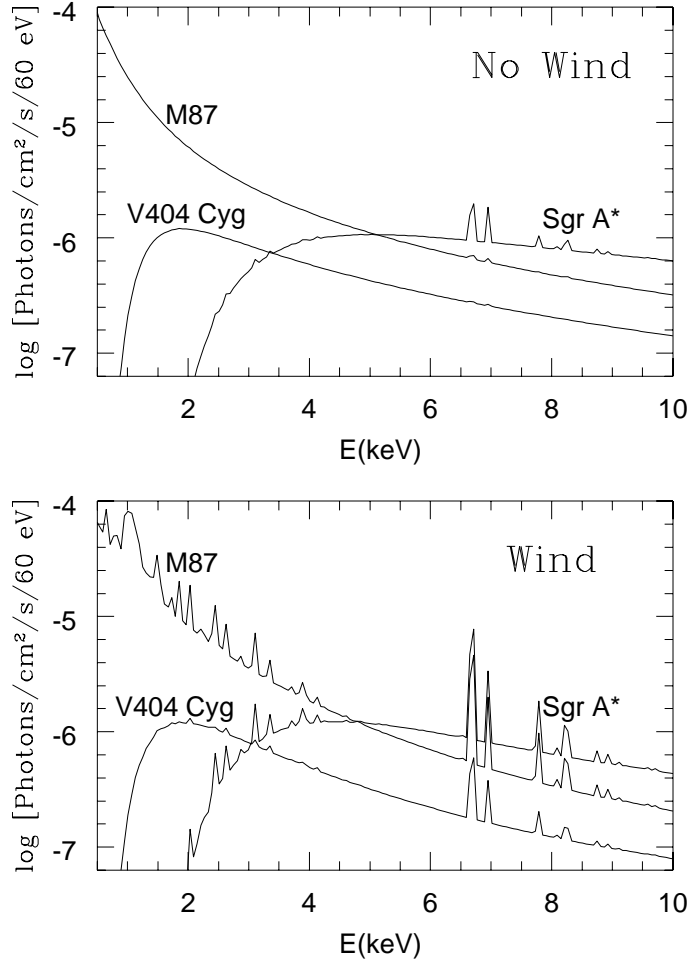}
\caption{Upper panel: Model X-ray line spectra of V404 Cyg (in
quiescence), Sgr A$^*$, and the nucleus of M87 (the spectrum of M87 is
shifted downward by 0.3 in the log for clarity).  The models assume
that the accretion occurs via an ADAF with no mass loss to a wind.
The spectra are shown with energy bins of 60 eV and have been
corrected for interstellar absorption with the appropriate values of
$N_H$.  Lower panel: Corresponding models for the case when there is a
moderately strong wind: $\dot m\propto r^{0.4}$.}
\end{figure}

\newpage
\vskip 5in
\newpage

\begin{figure}
\plotone{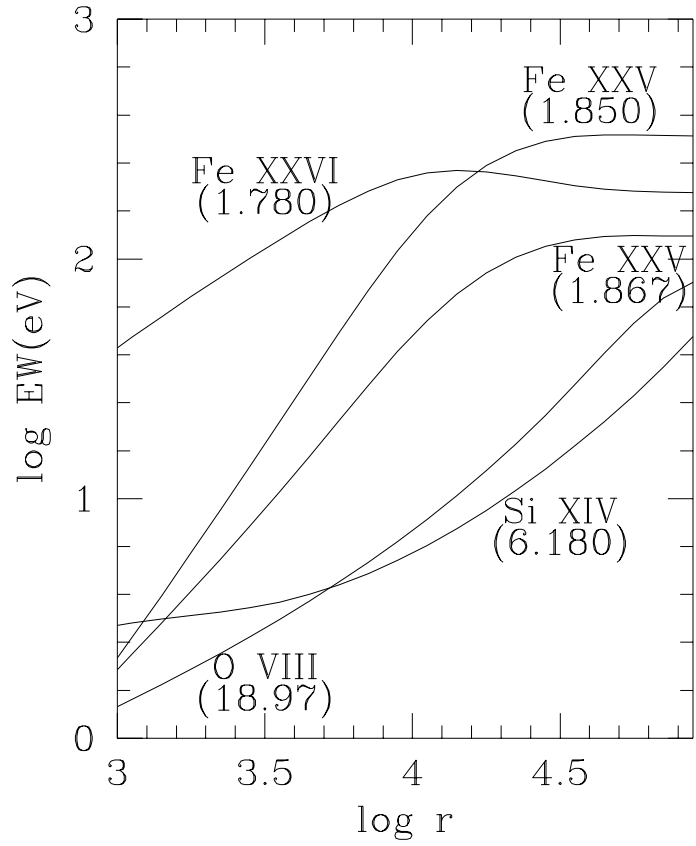}
\caption{Variation with radius $r$ of the equivalent widths of
selected X-ray lines (identified by wavelength in {\AA}) in the model
of Sgr A$^*$ with a wind.  Note that different lines saturate at
different radii, so that by studying relative line intensities it
should be possible to determine the amount of gas present at different
radii.}
\end{figure}

\end{document}